\documentclass[preprint,12pt]{elsarticle}

\usepackage{amssymb}

\begin{document}

\begin{frontmatter}



\title{Quantum interference through gated single-molecule junctions}


\author[imit]{Daniel A. Lovey}
\author[imit,facena]{Rodolfo H. Romero}
\ead{rhromero@exa.unne.edu.ar}
\address[imit]{Instituto de Modelado e Innovaci\'on Tecnol\'ogica, CONICET, Avenida  Libertad 5500 (3400) Corrientes, Argentina}
\address[facena]{Facultad de Ciencias Exactas y Naturales y Agrimensura, Universidad Nacional del Nordeste, 
Avenida  Libertad 5500 (3400) Corrientes, Argentina.}
\begin{abstract}
We discuss the general form of the transmission spectrum through a molecular junction in terms of the Green function of the isolated molecule. By introducing a tight binding method, we are able to translate the Green function properties into practical graphical rules for assessing beforehand the possible existence of antiresonances in an energy range for a given choice of connecting sites. The analysis is exemplified with a benzene molecule under a hypothetical local gate, which allows one to continuously tune the on-site energy of single atoms, for various connection topologies and gate positions.

\end{abstract}


\end{frontmatter}



\section{Introduction}
The theoretical prediction and experimental demonstration of the possibility of conduction through single molecules attached to metallic leads \cite{Aviram74, Aviram88, Samanta96, Reed97, Cui01}, has triggered the interest about the factors that affect the electron transmission and the feasibility to control them \cite{Stadler04, Xu05, Perrine07, Song09}. It is expected not only to scale down the electronic devices to smaller  sizes, but also to provide them with new functionalities based on the quantum nature of the system \cite{Stadler04, Hod08, Darau09, Stafford07}. In particular, the possibility of controling quantum interference has raised a great deal of interest from both a theoretical and experimental point of view \cite{Hod08, Patoux97, Baer02, Cardamone06, Kiguchi06, Kiguchi07, Ke08, Stadler09, Rincon09, Markussen10b, Rai11, Chen11}.
For instance, it has been pointed out the possibility of controling magnetically the transmission through molecular rings, such as aromatic molecules,  taking advantage of the condition of quantum interference \cite{Rai11}. 
Another way of affecting the quantum transmission could be by the direct modulation of transport current through a molecule junction by an external gate voltage \cite{Xu05, Song09}. Such a three-terminal molecule-scale device allows for the direct modification of the orbital energy. The physically interesting, although currently hypothetical, possibility of applying a local gate potential on one or more atoms in the molecule, would allow one to tune the molecular electronic structure in a similar way as the introduction of a substituent or side group does, although now in a continuously controlled way.

The standard first-principle theoretical description of the transmission through single molecule junctions is based on non equilibrium Green function (NEGF) techniques in conjunction with density functional theory (DFT) based electronic structure. Nevertheless, it has been shown that more approximate descriptions, such as Hartree-Fock or tight binding methods, also provide a qualitative and even reasonably quantitative description of the phenomena. Tight binding models are specially appealing, due to their simplicity, for providing insight into the most relevant physical mechanisms of the transmission as well as for using them as simple predicting tools for guiding either experiments or more precise theoretical methods.
In line with this, a number of methods have been proposed to discuss the relation between the electronic or molecular structure to the transmission properties of the molecular devices \cite{Yoshizawa08, Solomon08, Hansen09, Fowler09, Markussen10a, Dumont11, Ernzerhof11, Lovey11, Taniguchi11, Tsuji11}. 

In this work we study the quantum transmission through a molecular junction whose on-site energies can be continuously tuned by a local potential. As usual and for brevity, we shall refer here to such a potential as a gate potential. 
Although gating a single atom is currently only a hypothetical possibility, its theoretical analysis illustrates the physical principles underlying those effects. 
Firstly, we summarize the analytical properties of the transmission function in terms of the electronic structure and describe the two types of zeroes of resonance. Then, we use a tight binding model, to translate the analytical conditions on the Green function of the isolated molecule into graphical rules based on the molecular orbital plots. This graphical interpretation allows one to infer the complete spectrum of transmission, for an arbitrary connection, from a direct inspection of the molecular orbitals. Finally, we discuss how the application of a local potential at one or more atomic sites of the benzene molecule, gives rise to changes in the electronic structure which, in due course, modulates the quantum transmission through the molecule.
\section{\label{properties} Properties of the transmission function $T(E)$}
Consider that the atoms $l$ and $r$ of the molecule are connected to the sites $L$ and $R$ at the left and right leads by hopping parameters $t_L$ and $t_R$, respectively. Applying Dyson equation, the Green function ${\cal G}_{lr}$ of the coupled system can be obtained in terms of the Green functions of the isolated leads $G_{LL}$ and $G_{RR}$, and those of the isolated molecule $G_{ll}$, $G_{rr}$ and $G_{lr}$
\begin{equation}
{\cal G}_{lr} = \frac{G_{lr}}{1+\Sigma^2(G_{ll}G_{rr}-|G_{lr}|^2)-\Sigma(G_{ll}+G_{rr})},
\end{equation}
where $\Sigma=G_{LL}t_L^2=G_{RR}t_R^2$ is the self-energy, and symmetric coupling to identical leads is assumed. Within the wide band approximation, $\Sigma=i\Gamma$ is assumed to be purely imaginary and energy-independent, with $\Gamma$ giving the broadening of the molecular levels, such that
\begin{equation}
{\cal G}_{lr} = \frac{G_{lr}}{1-\Gamma^2(G_{ll}G_{rr}-|G_{lr}|^2)-i\Gamma(G_{ll}+G_{rr})},
\label{connected G1n}
\end{equation}
from which the transmission $T_{lr}$ can be obtained:
\begin{equation}
T_{lr}= 4\Gamma^2|{\cal G}_{lr}|^2.
\label{transmission}
\end{equation}

In approximate treatments of the electronic structure of molecules, like in the tight-binding (H\"uckel) model, the charge transport results from a competition between the on-site energy $\varepsilon$ that tends to localize the electron in the atom positions, and the hopping energy $t$ that favours the motion from an atom to its nearest neighbour.
As a result, the spectrum of transmission for a weakly coupled molecule ($\Gamma\ll t$) typically consists of narrow peaks of high (eventually perfect) conductance and narrower antiresonances (i.e., complete suppression of transmission) or dips at a discrete set of energy values, on top of a background of a smooth transmission function. This background has an approximately constant order of magnitude within an energy range, decreasing with a power-law of the energy out of it.
The origin of this smooth transmission function can be understood by neglecting the details of the molecular structure. Then, the molecule is characterized by a set of energy eigenvalues, roughly in the range $|E-\varepsilon|\lesssim 2t$, where the Green function behaves approximately as ${\cal G}_{lr}\sim 1/2t$, so the transmission becomes $T_{lr}=4\Gamma^2 |{\cal G}_{lr}|^2\sim \Gamma^2/t^2$.
It will be shown below that the peaks and dips reflect the details of the molecular junction, i.e., both the electronic structure of the molecule as well as the topology of the connection.

Eqs. (\ref{connected G1n}) and (\ref{transmission}) shows that the transmission through the connected molecule depends on the electronic structure of the isolated molecule both through the diagonal Green functions ($G_{ll}$ and $G_{rr}$) at the connecting sites as well as through the off-diagonal function $G_{lr}$ between them. The choice of the pair $(l,r)$ corresponds to the dependence on the topology of the connection.\\
\subsection{\label{resonance zeroes} Resonance and multipath zeroes}
Consider an energy eigenvalue $E_k$ of the disconnected molecule corresponding to the molecular orbital $|\psi_k\rangle$, which is written as a linear combination of the orbitals $|i\rangle$ centered at the atoms, $|\psi_k\rangle=\sum_i c_{ki} |i\rangle$, with $c_{ki}=\langle i| \psi_k \rangle$.
Due to the spectral representation of the Green function, the poles of $G$ are a set of (eventually all) the energy eigenvalues $E_k$.

It has been shown that, for non degenerate $E_k$, the poles of $G_{lr}$ are also poles of $G_{ll}$ and $G_{rr}$ \cite{Lovey11}. This entails that the Green function of the connected molecule, Eq. (\ref{connected G1n}), can be approximated near the pole $E=E_k$ as 
\begin{equation}
{\cal G}_{lr} \approx \frac{R_{lr}^k}{(E-E_k)-i\Gamma(R_{ll}^k+R_{rr}^k)} \stackrel{E\rightarrow E_k}{\longrightarrow} \frac{iR_{lr}^k}{\Gamma(R_{ll}^k+R_{rr}^k)},
\label{finite transmission}
\end{equation}
where $R_{ij}^k=c_{ki} c_{kj}^{*}$ ($i,j=l,r$) is the corresponding residue of $G_{ij}$, which shows that the transmission has a pole at $E=E_k+i\Gamma(R_{ll}+R_{rr})$ and a finite transmission $T_{lr}= 4R^2_{lr}/(R_{ll}^k+R_{rr}^k)^2$ at $E=E_k$. The level $E_k$ acquires a finite width proportional to the coupling to the leads $\Gamma$. In the particular case where the sites $l$ and $r$ are topologically equivalent because of the symmetry of the system, $R_{ll}=R_{rr}=R_{lr}$, a perfect transmission occurs.

On the other hand, if $E=E_k$ is a pole of $G_{ll}$ or $G_{rr}$, but not of $G_{lr}$, the numerator $G_{lr}(E_k)$ of Eq. (\ref{connected G1n}) is finite whilst its denominator diverges; therefore $T_{lr}$ will show an antiresonance at $E=E_k$.

Finally, the condition for the energy eigenvalue $E_k$ to be a pole of $G_{lr}$ can be related to the corresponding eigenstate $\psi_k$ of the isolated system: $E_k$ will become a pole of $G_{lr}$ if $\psi_k$ have non-vanishing projection on the orbitals $|l\rangle$ and $|r\rangle$.

Therefore, we can summarize the relation between the transmission and the electronic structure as follows: {\em the transmission coefficient $T(E)$ will show a peak of transmission, an antiresonance or a regular transmission $4\Gamma^2/E_k^2$ at the molecular energy $E_k$ if the molecular orbital $\psi_k$ has a non-vanishing weight at both, only one or none of the connecting sites, respectively.} 
\\ \indent

The case when the eigenvalue $E_k$ is degenerate requires some further consideration. For the sake of simplicity, consider a two-fold degenerate level $E_k$ with eigenfunctions $|\psi_k^{(p)}\rangle=\sum_i c_{ki}^{(p)} |i\rangle$, ($p=1,2$). The spectral representation of the Green function now reads
\begin{equation}
G_{ij}(E) = \sum_k \frac{R_{ij}^k }{E-E_k}, \hspace{0.5cm}
R_{ij}^k = c_{ik}^{(1)} c_{jk}^{(1)*} + c_{ik}^{(2)} c_{jk}^{(2)*}.
\label{spectral representation degenerate}
\end{equation}
The property $R_{ll}R_{rr}-R_{lr}^2= 0$, valid for the non-degenerate case, does no longer hold here. Instead
\begin{eqnarray}
R_{ll}^k R_{rr}^k-(R_{lr}^k)^2 &=& (c_{kl}^{(1)})^2 (c_{kr}^{(2)})^2 + (c_{kl}^{(2)})^2 (c_{kr}^{(1)})^2 \nonumber \\
                      &&- c_{kl}^{(1)} c_{kr}^{(1)} c_{kl}^{(2)*} c_{kr}^{(2)*} - c_{kl}^{(2)} c_{kr}^{(2)} c_{kl}^{(1)*} c_{kr}^{(1)*},\nonumber \\
\label{condition degenerate}
\end{eqnarray}
which can vanish or not depending on the $c_{kl}^{(p)}$ and $c_{kr}^{(p)}$. If $R_{ll}^k R_{rr}^k-(R_{lr}^k)^2 = 0$, all above discussion holds; if not,  the transmission becomes suppressed ($T=0$) \cite{Lovey11}.\\


The antiresonances discussed above occur at the energy eigenvalues of the molecule and, consequently, they were named as {\em resonant zeroes} of transmission.  Nevertheless, Eq. (\ref{connected G1n}), shows that also the roots of $G_{lr}$ can lead to zeroes of conductance, that were termed {\em multipath zeroes} \cite{Hansen09}. To explore their origin, we employ a tight binding Hamiltonian together with a partitioning consisting in dividing the basis of orbitals centered at the sites in the $P$-subspace ($\{|l\rangle,|r\rangle\}$) and the $Q$-subspace, formed by the rest of the orbitals centered at non-connecting sites.

The Green function of the isolated molecule can be written as the resolvent $(E-H_{\rm eff})^{-1}$ of an effective $2\times 2$ Hamiltonian in the $P$-space, in terms of the self-energy $\Sigma(E)$ as \cite{Hansen09, Lovey11}
\begin{equation}
G(E) = \frac{1}{\Delta} \left( 
\begin{array}{cc}
E-\varepsilon_r-\Sigma_{rr} & t_{lr}+\Sigma_{lr} \\
t_{rl}+\Sigma_{rl} & E-\varepsilon_l-\Sigma_{ll}
\end{array}
\right),
\end{equation}
where $\Delta(E)=\det(E-H_{\rm eff})=(E-\varepsilon_r-\Sigma_{rr})(E-\varepsilon_l-\Sigma_{ll})-|t_{lr}+\Sigma_{lr}|^2$. The energy-dependent diagonal and off-diagonal elements of the self-energy are interpreted here as effective on-site energies and effective hoppings between the sites $l$ and $r$, respectively. Now assume that sites $l$ and $r$ are coupled to the rest of the molecule through sites $l'$, $l''$, $r'$ and $r''$, as shown in Fig. 1, through identical hoppings. \\
%
\begin{figure}
\begin{center}
\includegraphics[scale=0.2]{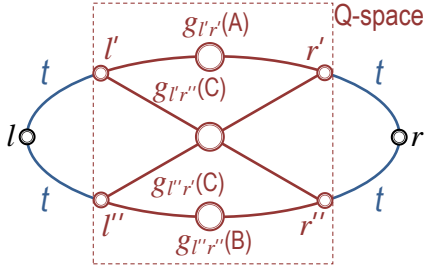}
\end{center}
\caption{\label{Q-space scheme} Scheme of the partitioning of the Hamiltonian. The leads are attached to sites $l$ and $r$, which are connected to the rest of the molecule through atoms $l'$, $l''$, $r'$ and $r''$. The Green function $G_{lr}$ results from hoppings $t$ from $l$ to the $Q$-space, a sum of contributions $g_{ij}$ along pathways (A, B and C) within the $Q$-space, and hoppings $t$ from the $Q$-space to the atom $r$. The large empty circles represent atomic sites that are not directly connected to the atoms $l$ or $r$.}
\end{figure}
Then, 
\begin{equation}
G_{lr}=\frac{t_{lr}+\Sigma_{lr}}{\Delta},
\end{equation}
where $t_{lr}=t$ if $l$ and $r$ are adjacent to each other, and zero otherwise, with 
\begin{equation}
\Sigma_{lr}= (g_{l'r'}+g_{l''r''}+g_{l'r''}+g_{l''r'})t^2,
\end{equation}
where $g$ refers to the $Q$-block of the Green function $G$. This gives $G_{lr}$ as a sum of four terms that can be interpreted as the pathways A (above), B (below) and C (cross) shown in Fig. 1, and the Green function has the form
\begin{equation}
g = \left( \begin{array}{cc}
g^A & g^C \\ g^{C\dag} & g^B
\end{array}\right).
\end{equation}

When the molecule has two paths disconnected from each other, i.e., two sets of disjoint sites, as in a cyclic molecule, the Green function in the $Q$-space becomes blocked
\begin{equation}
g = \left( \begin{array}{cc}
g^A & 0 \\ 0 & g^B
\end{array}\right),
\end{equation}
so that, $g_{l'r''}=g_{l''r'}=0$, and there are only two contributing pathways A ($l'\leftrightarrow r'$) and B ($l''\leftrightarrow r''$):
\begin{equation}
\Sigma_{lr} = (g_{l'r'}+g_{l''r''})t^2.
\label{2 pathways}
\end{equation}
Then, a zero of transmission coming from $G_{lr}=0$ can be attributed to a cancellation of the two contributions, one from each path: $g_{l'r'}+g_{l''r''}=0$. 
\subsection{\label{multipath zeroes} Approximate positions of the multipath zeroes}
From the spectral representation it can be seen that near a pole $E_n$ the Green function behaves like $G_{lr}(E) \approx c_{nl}c_{nr}/(E-E_n)$, where $|c_{ni}|\le 1$ and, therefore, for $E_n\le E \le E_{n+1}$, its behaviour is well represented by 
\begin{equation}
G_{lr}(E) = \frac{c_{nl}c_{nr}}{E-E_n} + \frac{c_{n+1,l}c_{n+1,r}}{E-E_{n+1}},
\label{approx green}
\end{equation}
which is schematically depicted in Figs. 2(a)-2(b) for $c_{nl}c_{nr}>0$, and positive (a) or negative (b) $c_{n+1,l}c_{n+1,r}$.\\
\begin{figure}
\vspace{-1cm}
\includegraphics[scale=0.5]{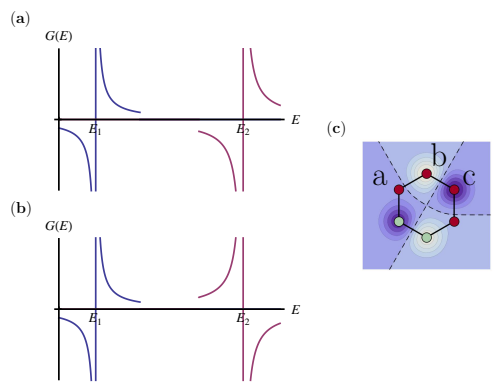}
\vspace{-7.5cm}
\caption{\label{GFs scheme} Scheme of the Green function $G(E)$ near two consecutive poles $E_1$ and $E_2$ and the visualization of their significance in the MO plot: Residues at $E=E_1$ and $E=E_2$ have (a) the same sign, (b) opposite sign. (c) Plot of a typical molecular orbital (MO): nodal (dashed) lines in the plane of the molecule separate darker and lighter regions of positive and negative sign of the wave function; the closer an atom to the nodal line, the smaller its coefficient in the MO expansion in terms of atomic orbitals. Pairs of atoms placed in regions of opposite sign (e.g., atoms a and b) give negative residues while those in regions of the same sign (e.g., a and c) give positive residues in the corresponding Green functions at the energy of the MO. Furthermore, $|{\rm Res} (g_{\rm ab})|< |{\rm Res} (g_{\rm bc})|$ because atom `a' is close to a nodal line.}
\end{figure}
When both residues have the same sign, say positive, $g(E)$ changes its sign from $g(E_n+0)=\infty$ to $g(E_{n+1}-0)=-\infty$, thus showing that there exists a root in the range $(E_n,E_{n+1})$ which in due turn produces a zero of resonance at a weighted average of $E_n$ and $E_{n+1}$
\begin{equation}
E = \frac{c_{nl}c_{nr}E_{n+1}+c_{n+1,l}c_{n+1,r} E_n}{c_{nl}c_{nr}+c_{n+1,l}c_{n+1,r}}.
\label{root position}
\end{equation}
It should be noted that $E_n$ is weighted by the residue at $E_{n+1}$ such that the larger the residue is at $E_{n+1}$, the closer is the root to $E_{n}$. This could be pictorially described as a repulsion of the root by two successive poles that enclose it and having residues of the same sign.

On the other hand, if both residues have opposite signs,
\begin{equation}
G_{lr}(E) = \frac{c_{nl}c_{nr}}{E-E_n} - \frac{|c_{n+1,l}c_{n+1,r}|}{E-E_{n+1}}>0,
\end{equation}
becomes definite positive in the range $(E_n,E_{n+1})$ and no zero occurs. Rather rarely, the approximation (\ref{approx green}) might not be accurate enough, because other poles make non negligible contributions, so that the curve cuts the $E$-axis twice and two antiresonances occurs.
\subsection{Predicting transmission spectra from orbital plots}
Now we shall show that the general form of the transmission spectra can be obtained from a graphical interpretation of the properties derived in Section \ref{properties} from the plots of the molecular orbitals. Typical plots show regions of different signs (shown as darker and lighter zones) separated by nodal lines resulting from the intersection between nodal surfaces with the plane of the molecule (see Fig. 2c) The following observations translate the analytical properties of the coefficients $c_{ki}=\langle i|\psi_k \rangle$ into graphical rules:

({\em i}) If a molecular orbital $\psi_k$ has a vanishing weight at any of the sites of connection, a nodal line in the plot of $\psi_k$ passes through it. 

({\em ii}) When the atoms connected to the leads ($l$ and $r$) are in regions of opposite sign of $\psi_k$, it entails a negative residue ($c_{kl}c_{kr}< 0$) at $E=E_k$; analogously, when the atoms $l$ and $r$ are in regions of the plot where $\psi_k$ has the same sign, the residue of the pole $E=E_k$ has a positive sign. 

({\em iii}) Due to $\psi_k$ is normalized, the proximity of the atom $l$ or $r$ to a nodal line entails that the coefficient is small, and so is the residue at $E=E_k$.\\

The molecular orbital plots can also be directly related to the transmission spectra as follows:

(I) In the transmission spectrum, there are peaks and antiresonances at the molecular eigenvalues $E_k$ . The peaks occurs at those values $E_k$ whose corresponding molecular orbital has no nodal line passing through the connection sites, the antiresonances when it does.

(II) The transmission spectrum can have additional antiresonances at energies between two consecutive molecular energy eigenvalues $E_n$ and $E_{n+1}$ of the molecular orbitals $\psi_n$ and $\psi_{n+1}$ (Section \ref{multipath zeroes}). 
If in the plots of $\psi_n$ and $\psi_{n+1}$, the two atoms $l$ and $r$ have the same relative phase (both atoms in regions of the same sign or both in regions of opposite signs), there will be an antiresonance at $E$ given by Eq. (\ref{root position}).

(III) The closer the atom $l$ or $r$ is to a nodal line, the smaller the transmission for the connection $(l,r)$.

The simple rules stated above allows us to explain the calculated transmission spectra shown in the next Section.
\section{Results and discussion}
As an example of the analysis given in the previous section, we calculated the transmission through benzene in its {\em ortho}-, {\em meta}- and {\em para}-connections to the leads. Fig. 3 shows the scheme of the molecular orbitals for benzene (disconnected from the leads) with a gate voltage of 0.5 eV  applied on atom 5, {\em i.e.}, in the later case, the on-site energy $\varepsilon_5$ is shifted to $\varepsilon_g=-6.1$ eV while the rest of the atoms are kept at $\varepsilon=-6.6$ eV. The hopping parameter is $t=-2.5$ eV, and the coupling to the leads $\Gamma=0.05$ eV.

 The application of the gate potential produces small changes in the energy eigenvalues with respect to the ungated molecule. Nevertheless, two qualitatively different features occur, namely, the breaking of the degeneracy of states $(E_2,E_3)$ and $(E_4,E_5)$, and the change in the symmetry of the corresponding orbitals. The lowest and the highest energy orbitals are only slightly modified and keep their symmetry. For both, gated and ungated benzene, the first orbital have a single sign in all the plane while the highest energy orbital alternates its sign from one atom to the adjacent one. \\
%
\begin{figure}
\begin{center}
\vspace{-2cm}
\hspace{-2.3cm}
\includegraphics[scale=0.6]{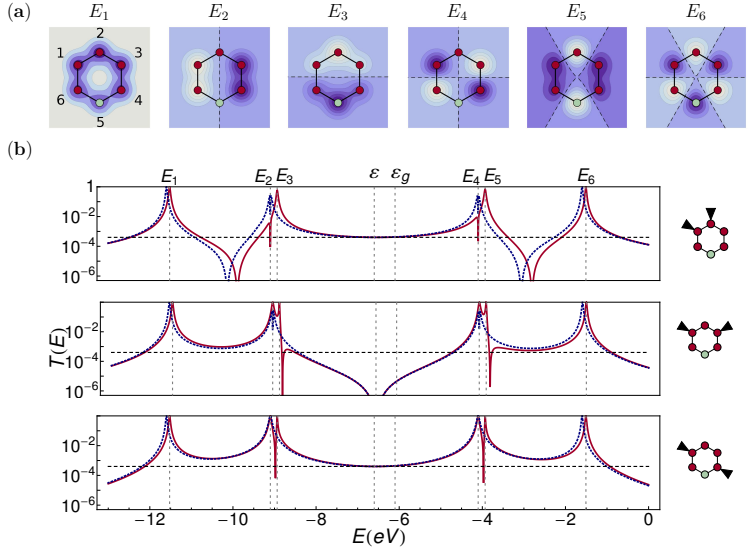}
\end{center}
\vspace{-7.5cm}
\caption{\label{gate on 5} Transmission function $T(E)$ for: (a) 1,2 {\em ortho}-benzene, (b) 1,3 {\em meta}-benzene and (c) 1,4 {\em para}-benzene with a gate potential applied at atom 5 (solid red line). The dashed blue lines are the transmission through ungated benzene for comparison. The horizontal dotted black lines are the order of magnitude estimates $\Gamma^2/t^2=4\times 10^{-4}$.} 
\end{figure}

From application of the above stated graphical rules, the effect of the relative positions of gate and leads can be anticipated. Firstly, we consider the transmission spectra of benzene with a gate at atom 5. The left lead is connected at atom $l=1$, while the right lead is attached to atom $r$, where $r=2$, 3 and 4 for the {\em ortho}, {\em meta} and {\em para} connections, respectively.
Fig. 3a shows plots of molecular orbitals for this case.  For none of them any nodal line passes through atoms 1, 3 nor 4. Therefore, at the energy eigenvalues there are peaks of transmission in the {\em meta} and {\em para} connections. For the {\em ortho} connection, however, there are zeroes of transmission at $E=E_2$ and $E=E_4$ because a nodal line passes through atom 2 in the plot of the molecular orbitals corresponding to those energies. The other two antiresonances are multipath zeroes in the intervals $(E_1,E_2)$ and $(E_5,E_6)$, that are due to the fact that the atoms 1 and 2 have residues of the same sign at two consecutive poles of the Green function. In the plots, this is seen as due to the fact that the atoms 1 and 2 both occupy regions of opposite signs in the two consecutive orbitals $\psi_5$ and  $\psi_6$.

On the other hand, atoms 1 and 3 occupy regions of the same sign, thus giving positive residue, in the orbitals $\psi_1$, $\psi_3$, $\psi_5$ and $\psi_6$. Therefore an antiresonance is expected for the {\em meta} connection (with the gate applied at atom 5) in the interval $(E_5,E_6)$.

Finally, atoms 1 and 4 occupy regions of the same sign for the orbitals $\psi_1$, $\psi_4$ and $\psi_5$; as a consequence an antiresonance in the transmission function of the {\em para}-connection arises in the interval $(E_4,E_5)$, since both energies are poles with residues of the positive sign. A similar analysis shows that $E_2$ and $E_3$ are poles having both negative residues, thus giving an antiresonance in the interval $(E_2,E_3)$.\\

Fig. 4b shows the comparison for a fixed connection ({\em meta} between atoms 1 and 3) of the effect of the gate position. The transmission functions for a gate applied on atoms 4 and 5 are shown with dashed blue line and solid red line, respectively. The most remarkable change is that the peak of transmission at $E=E_2$ and $E=E_4$, when the gate is applied on atom 4, turns into an antiresonance when the gate is applied on atom 5. The reason can be traced back to the Green function (depicted in Fig. 4c), which shows that $G_{13}(E)$ diverges at $E=E_2$ for the gate at 5, while remains finite for the gate at 4. Nevertheless, a simpler justification can be argued based on the molecular orbital plots. As seen in Fig. 4a, both sets of orbitals only differ in a 60 degrees rotation between each other. Then, the nodal line passing through atom 1 at $E=E_2$ and $E=E_4$ when the gate is on site 4, transforms into a nodal line passing through atom 2 when the gate is on site 5. Then, for the 1,3-{\em meta}- connection, the former has a vanishing weight at one of the connecting atoms (atom 1), while the latter does not.
Interestingly, such a change in the relative position of gate and connecting sites, entails a change from perfect to null transmission at a given energy.\\

In order to get a deeper insight of the cancellation of the transmission due to  interference of pathways, consider benzene connected to the leads in the {\em meta} configuration with $l=1$, $r= 3$, and the gate potential applied to atom 4 or 5. The self-energy $\Sigma_{lr}=\Sigma_{13}$ becomes the sum of two contributions [Eq. (\ref{2 pathways})]
\begin{equation}
\Sigma_{13} = (g_{22}+g_{46})t^2,
\end{equation}
one through the single site 2 with $g_{22}=(E-\varepsilon)^{-1}$, and another through a three-atoms chain ($4 \rightarrow 5 \rightarrow 6$) having an on-site energy $\varepsilon_g$ at its end (gate applied on site 4) or in the central site (gate applied at 5). Then, 
\begin{equation}
g_{46}^{(4)} = \frac{t^2}{(E-\varepsilon)^2(E-\varepsilon_g)+t^2(\varepsilon_g+\varepsilon-2E)} , (\varepsilon_g {\rm \ on \  site \ 4}),
\end{equation}
or 
\begin{equation}
g_{46}^{(5)} = \frac{t^2}{(E-\varepsilon)\left[(E-\varepsilon)(E-\varepsilon_g)-2t^2\right]} , (\varepsilon_g {\rm \ on \ site \ 5})
\end{equation}
with the super-index refering to the gate position. 

When the gate is applied on site 4, at the energy $E=\varepsilon_g$, the contributions $g_{46}^{(4)} = (\varepsilon-\varepsilon_g)^{-1}=-g_{22}$, cancels each other exactly, thus giving $\Sigma_{13}=0$ and therefore, an antiresonance in the transmission occurs at the energy of the gate potential.

On the other hand, if the gate potential is applied at site 5, both $g_{22}$ and $g_{46}^{(5)}$ diverge at $E=\varepsilon$, and so does $\Sigma_{13}$. 
Nevertheless, for this analytically solvable case, it can be explicitly shown that for arbitrary on-site energies  $\varepsilon_4$ and $\varepsilon_5$, the self-energy is proportional to $(E-\varepsilon_4)\left[(E-\varepsilon_5)(E-\varepsilon)-t^2\right]$. Therefore, when the gate is appled on site 5, $\varepsilon_4=\varepsilon$ and $\varepsilon_5=\varepsilon_g$, the antiresonances occur at $E=\varepsilon$ and $(\varepsilon+\varepsilon_g)/2\pm t$, which correspond to the zeroes at -8.85 eV and -3.85 eV.\\

Consider now {\em para}-connected benzene: $l=1$, $r= 4$; in this case, $\Sigma_{14}=(g_{23}+g_{56})t^2$. If all on-site energies  $\varepsilon$ are the same, $g_{23}=g_{56}$ and there is constructive interference. However, if one on-site energy, say $\varepsilon_2=\varepsilon_g$, is different,
\begin{equation}
g_{23} = \frac{t}{(E-\varepsilon_a)(E-\varepsilon_b)}, \hspace{0.5cm}
g_{56} = \frac{t}{(E-\varepsilon'_a)(E-\varepsilon'_b)}, 
\end{equation}
where $\varepsilon_a$ and $\varepsilon'_a$ are the energies of the antibonding orbitals, while $\varepsilon_b$ and $\varepsilon'_b$ are those of the bonding ones for the fragments 2-3 and 5-6, respectively,
\begin{eqnarray}
\varepsilon_{a,b} &=& 
(\varepsilon_g+\varepsilon)/2\pm\sqrt{t^2+(\varepsilon_g -\varepsilon)^2/4}, \\
\varepsilon'_{a,b} &=& 
\varepsilon\pm t 
\end{eqnarray}
Therefore, for energies $\varepsilon_a < E < \varepsilon'_a$ and $\varepsilon_b < E < \varepsilon'_b$, $g_{23}$ and $g_{56}$ have opposite signs and can cancel each other, thus giving a vanishing transmission. It should be noted that no antiresonance occur in the interval $(\varepsilon'_a,\varepsilon_b)$, roughly corresponding to the absence of zeroes of transmission within the HOMO-LUMO gap in the {\em para}-connection.
Hence, the two paths interfere destructively as a consequence of shifting the bonding and anti-bonding levels of one fragment with respect to the other, by tuning one on-site energy.\\

\begin{figure}
\vspace{-1.5cm}
\begin{center}
\hspace{-2.1cm}
\includegraphics[scale=0.5]{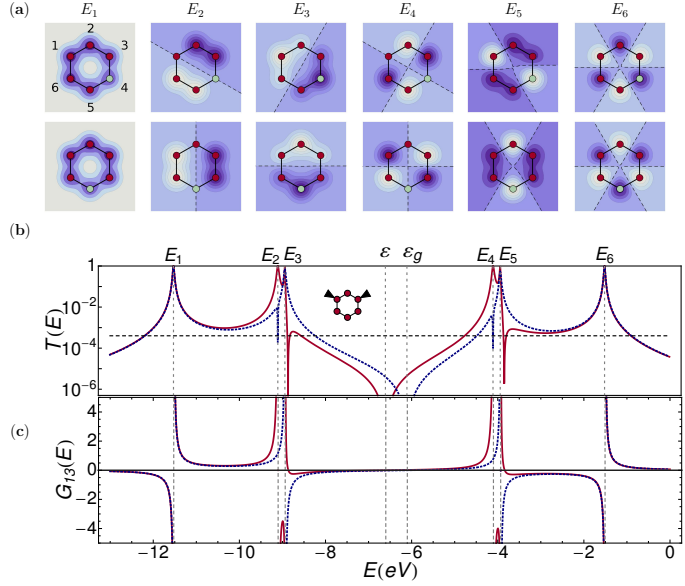}
\end{center}
\vspace{-5cm}
\caption{\label{compare gate on 4-5} (a) Plots of molecular orbitals for the gate potential applied on atom 4 and 5. (b) Transmission function $T(E)$ and (c) the Green function for 1,3 {\em meta}-benzene with gate potentials applied at atoms 4 (dashed blue line) and 5 (solid red line).} 
\end{figure}
Finally, Fig. 5 shows the effect of applying a gate voltage on atoms 5 and 6 simultaneously. From applying the graphical rules to the plots of the molecular orbitals, Fig. 5(a), the transmission in the {\em ortho}, {\em meta} and {\em para}-connections, Fig 5(b)-(d), can be understood as follows. 
Since none nodal surface passes through any atom, there are peaks of transmission at every molecular eigenenergy. Nevertheless, two antiresonances arise close to energies $E_3$ and $E_5$, as shown in the insets. The orbital $\psi_4$ shows that, at the energy $E=E_4$, each one of the connected atoms is far from the nodal lines. On the contrary, the orbital $\psi_5$ shows that the atom 1, connected to the left lead, is very close to a nodal surface, thus implying that the weight of the orbital on atom 1 is small. The coefficients of $\psi_5$ on atoms 2 and 3 are larger, thus producing a finite, although not perfect, transmission for the {\em ortho} and {\em meta} connections. In the {\em para} connection, nevertheless, both atoms 1 and 4 are equivalent because their positions are symmetrical with respect to an axis passing perpendicularly to the line joining the gated atoms 5 and 6, thus producing a perfect transmission, as mentioned for Eq. (\ref{finite transmission}).\\
%
\begin{figure}
\begin{center}
\hspace{-1.5cm}
\includegraphics[scale=0.6]{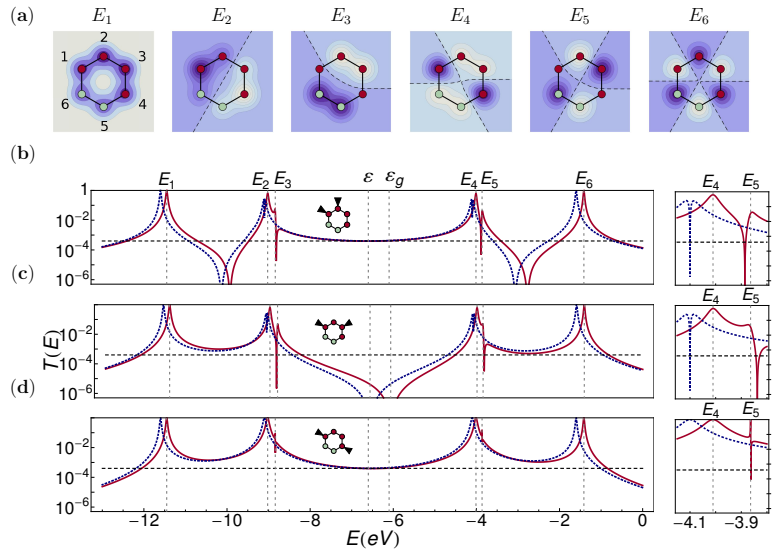}
\end{center}
\vspace{-7.7cm}
\caption{\label{gates on 5 and 6} (a) Scheme of the molecular orbitals with gate potentials applied at atoms 5 and 6 shown in (green) filled circles, and transmission function $T(E)$ (shown in solid red lines) for: (b) 1,2 {\em ortho}-benzene, (c) 1,3 {\em meta}-benzene and (d) 1,4 {\em para}-benzene. The  transmission through ungated benzene are shown in dashed blue lines for comparison. The panels to the right show $T(E)$ in more detail around the energies $E_4$ and $E_5$.} 
\end{figure}
\section{Concluding remarks}
In this work we have analyzed the relation between the transmission function with the electronic structure of the isolated molecule and provided graphical rules of analysis from direct inspection of the plots of molecular orbitals.
As an application of the concepts discussed, we have calculated the transmission of a benzene molecule subjected to tunable gate potentials, represented by a continuous variation of the on-site energy with the molecule connected in ortho, meta and para configurations. 
It is expected that this type of pictorial prediction of the transmission coefficient should be of help as a guideline for experiments and applications of more accurate theoretical methods.
\section*{Acknowledgements} This work was partly supported by SGCyT (Universidad Nacional del Nordeste), National Agency ANPCYT and CONICET (Argentina) under grants PI 112/07, PICTO-UNNE 204/07 and PIP 11220090100654/2010. 
\end{document}